# Half-integer Shapiro Steps in Strong Ferromagnetic Josephson Junctions

Yunyan Yao[1], Ranran Cai[1], See-Hun Yang[2], Wenyu Xing[1], Yang Ma[1], Michiyasu Mori[3], Yuan Ji[1], Sadamichi Maekawa[3,4,5], Xin-Cheng Xie[1,6,7], and Wei Han[1*]

[1] International Center for Quantum Materials, School of Physics, Peking University, Beijing 100871, P. R. China

[2] IBM Research - Almaden, San Jose, California 95120, USA

[3] Advanced Science Research Center, Japan Atomic Energy Agency, Tokai 319-1195, Japan

[4] RIKEN Center for Emergent Matter Science (CEMS), Wako 351-0198, Japan

[5] Kavli Institute for Theoretical Sciences (KITS), University of Chinese Academy of Sciences, Beijing 100049, P. R. China

[6] CAS Center for Excellence in Topological Quantum Computation, University of Chinese Academy of Sciences, Beijing 100190, P. R. China

[7] Beijing Academy of Quantum Information Sciences, Beijing 100193, P. R. China

* Correspondence to: weihan@pku.edu.cn.

**Abstract:**

We report the experimental observation of half-integer Shapiro steps in the strong ferromagnetic Josephson junction (Nb-NiFe-Nb) by investigating the current-phase relation under radio-frequency microwave excitation. The half-integer Shapiro steps are robust in a wide temperature range from $T$ = 4 to 7 K. The half-integer Shapiro steps could be attributed to co-existence of 0- and $\pi$-states in the strong ferromagnetic NiFe Josephson junctions with the spatial variation of the NiFe thickness. This scenario is also supported by the high-resolution transmission electron microscopy characterization of the Nb/NiFe/Nb junction.

I. INTRODUCTION

The AC Josephson effect has attracted a lot of attentions recently for the development of superconducting qubit-based quantum computing [1,2]. As a consequence of the AC Josephson effect, the Shapiro voltage steps could be induced in the presence of microwave excitation, which can be described by $V = nhf/2e$ ($V$ is the voltage, $n$ is the index of Shapiro steps, $h$ is the Planck constant, $f$ is the microwave frequency, and $e$ is the electron charge) [3]. Since it is due to the phase matching mediated by the irradiated microwave field and the driven electric field, the Shapiro steps are also called the phase-locking effect. The Shapiro steps could manifest the current-phase relation (CPR) of the Josephson junctions (JJs) [4]. And very interestingly, half-integer Shapiro steps could be observed in certain circumstances due to unconventional physical properties. For example, the half-integer Shapiro steps have been observed in the metals, semiconductors, and topological materials based JJs due to the skewed subharmonic CPR [5], the bound states dynamics [6,7], the intrinsic second harmonic CPR [8-10], and the Josephson interferometry effect [11,12].

Beyond these nonmagnetic JJs, there are ferromagnetic (FM) JJs which have been considered to be building blocks for quantum-flux qubits [13]. Besides, the FM JJs could be unique probes for long-range spin-triplet supercurrent [14-17] and the interplay between ferromagnetism and superconductivity [18,19]. Recently, half-integer Shapiro steps have been reported in the weak FM JJs, such as the CuNi alloy (exchange energy $E_{ex}$ ~ 5 meV) [8,9,11]. In the CuNi JJs, there is a transition between 0- and $\pi$- states (0- and $\pi$- states represent the Josephson coupling with the negative and positive sign, respectively [18]) as the temperature changes across the critical point, and the half-integer Shapiro steps were reported in the small temperature range of ~ 100 mK. Different from weak FMs, strong FMs, such as Co, Ni, and Fe, have large exchange energy over 100 meV [20]. Thus, the 0- and π- states transition is not able to realize by varying temperature. Alternatively, 0- and π- states transitions have been achieved by varying FM thickness [21,22], due to the oscillating decay of the superconducting wave function into the FM. However, the systematic investigation of the half-integer Shapiro steps in strong FM JJs has not been reported.

In this paper, we report the experimental observation of the half-integer Shapiro steps of a strong FM JJ made of Nb/Ni$_{80}$Fe$_{20}$ (NiFe)/Nb. The half-integer Shapiro steps are robust in a wide temperature range from 4 to 7 K, which might be useful for future FM JJs-based quantum qubit applications. We attribute the half-integer Shapiro steps to the co-existence of the 0- and π-states junctions, which arises from the spatial variation of the NiFe thickness in the Nb/NiFe/Nb heterostructures.

## II. EXPERIMENTAL

The Josephson junctions were made of Nb/NiFe/Nb trilayer heterostructures grown on thermally oxidized Si substrates in a d.c. magnetron sputtering system with a base pressure of ~1×10$^{-8}$ torr. The NiFe Josephson devices, illustrated in Fig. 1(a), were fabricated using the standard shadow mask technique. During the growth of each layer, several shadow masks were used to pattern the bottom Nb electrode (100 nm), NiFe layer (5.5 nm), Al$_2$O$_3$ layer (100 nm), and the top Nb electrode (100 nm), thus forming the NiFe JJ device. The Al$_2$O$_3$ layer was used as an electrical insulating layer. After the growth of these layers, a 2.3-nm Al$_2$O$_3$ capping layer was used to prevent *ex situ* oxidation/degradation of the junctions. The junction area is ~ 80 × 80 μm$^2$. The electrical measurement of the Josephson coupling was performed in the variable temperature insert of a Physical Properties Measurement System (PPMS; Quantum Design). The AC Josephson effect was measured by standard a.c. lock-in technique at low frequency of 7 Hz. The current was provided using the Agilent K6221 current source, consisting of a small AC current modulation (dI = 5 µA) and a DC current bias ($I_{DC}$). The differential voltage (dV) was detected using a lock-in amplifier (SR830). The microwave excitation was applied via the coplanar wave-guide technique connected with a vector network analyzer (VNA; Agilent E5071C) with the frequency ranging from 300 KHz to 20 GHz. Both the integer and half-integer steps can be clearly observed with microwave power from -5 to 5 dBm measured from the source output (Fig. S1 in the Supplemental Material [23]). The data shown in this paper were obtained under the radiated microwave power of 0 dBm unless noted. The half-integer Shapiro steps can be observed without and with magnetic fields (Fig. S2(a) in the Supplemental Material [23]), indicating the robustness of the half-integer Shapiro steps. A finite magnetic field is helpful to obtain clearer half-integer steps.

## III. RESULTS

Figure 1(b) shows the normalized differential resistances measured as a function of the current bias on the NiFe JJ device at various temperatures. Clearly, strong Josephson coupling is observed across the NiFe layer below $T = 7$ K. The critical current density ($J_c$) is defined as the point where the differential resistance increases above the value for the zero-bias current. As the temperature decreases, the critical current density increases and starts to saturate (Fig. 1(c)). The temperature dependence of the critical current is similar to previous reports on strong FM Josephson junctions [20]. Figure 1(d) shows the current-voltage character ristics of the NiFe JJ with microwave excitation at $T = 5$ K and under the in-plane magnetic field of 750 Oe. Clearly, there are voltage plateaus at around $\pm 20$ µV (see the dashed lines), indicating the possible appearance of the Shapiro steps [3,8,24]. For comparison, without microwave excitation, the voltage plateaus disappear (black curve in Fig. 1(d)). To check whether the voltage plateaus are Shapiro steps, the voltage is normalized by $\Phi f$ (where $\Phi = h/2e$ is the quantum flux), as shown in the right *y*-axis of Fig. 1(d). The dashed lines represent the n = 1 and n = -1 for the V/$\Phi f$, which matches the voltage plateaus well.

To further investigate the Shapiro steps in the strong FM JJs, we study the differential resistance at various RF frequencies at *T* = 5 K, as shown in Fig. 2(a). The RF frequency is varied in range from 8 to 12 GHz and the in-plane magnetic field is 750 Oe. The dark-blue area represents the superconducting state and the edges correspond to the critical current densities. Above the critical current densities, there are two clear blue stripes, which can be confirmed to be the integer Shapiro steps as indicated by the black arrows and labels of "$\pm 1$". Interestingly, besides these two stripes, there are obvious light-yellow lines that correspond to the half-integer Shapiro steps. To see these features more clearly, the differential resistance curves are plotted as a function of the voltage bias for several representative frequencies, as shown in Fig. 2(b). The dashed lines are guides to the eye for the half-integer and integer Shapiro steps. The numerical values of the voltage bias for these Shapiro steps are extracted as a function of the applied RF frequency, as shown in the Fig. 2(c). Based on the best fitting curves using the equation of $V = |n\Phi f|$ (n is the index of the Shapiro steps), the results of |n| are 0.508 $\pm$ 0.006, 1.01 $\pm$ 0.001, 1.53 $\pm$ 0.004, and 2.02 $\pm$ 0.01, respectively. Clearly, the linear relationship with half-integer and integer numbers confirms that the observed voltage dips are related to the half-integer and integer Shapiro steps.

Fig. 3(a) shows the contour mapping of the differential resistance as a function of microwave frequency (4 – 18 GHz) at *T* = 7 K under the in-plane magnetic field of 2300 Oe. The obvious "fan-like" behaviors are undoubtedly related to the frequency-dependent Shapiro steps indicated by the dashed lines. The integer and half-integer steps cannot be clearly resolved below 3 GHz. In Fig. 3(b), the differential resistance is plotted as a function of the normalized voltages under several RF frequencies, where the half-integer and integer Shapiro steps can be clearly resolved. When the temperature is higher than 7 K, the Josephson coupling is not observed, thus leading to the disappearance of Shapiro steps. The temperature range of the half-integer Shapiro steps (from $T = 4$ to 7 K) is much larger than that observed in the weak FM based JJs. For example, the half-integer Shapiro steps appear in the small temperature window of ~100 mK for CuNi JJs [11].

Next, the temperature dependence of the half-integer Shapiro steps in the strong FM Josephson junctions is investigated. To investigate the temperature dependence of half-integer steps, the results at 2300 Oe are shown in Fig. 4, since the half-integer steps at all the temperatures can be clearly observed. Figures 4(a-d) show the differential resistance (black curves) as a function of the current density for $f = 15$ GHz at $T = 4, 5, 6,$ and 7 K, respectively. And the corresponding current-voltage (normalized by $\Phi f$) curves (red lines) are also shown in Figs. 4(a-d), where the index of the Shapiro steps can be seen clearly. To characterize these Shapiro steps, the half-width ($\Delta J_n$) of Shapiro steps are investigated, and the $\Delta J_{-1}$ and $\Delta J_{-1/2}$ in Fig. 4(a) represent the half-width of the n = -1 and n = -1/2 Shapiro steps. The half-width of the integer Shapiro steps ($\Delta J_{|\pm 1|}$) decreases as the temperature increases from 4 to 7 K (Fig. 4(e)), which is strongly related to the critical current density expected by the resistively shunted junction (RSJ) model [25,26]. As temperature decreases, the critical current increases, leading to the enhancement of step width of integer Shapiro steps [27]. Different from the integer Shapiro steps, the half-width of the half-integer Shapiro steps ($\Delta J_{|\pm 1/2|}$) exhibits little variation as the temperature varying. Those temperature dependence of $\Delta J_{|\pm 1|}$ and $\Delta J_{|\pm 1/2|}$ have been also studied under the RF microwave with frequency of 8 GHz (Fig. 4(f)). Similar features are also observed in the magnetic field dependence of the step sizes for half-integer and integer Shapiro steps (Fig. S2(b) in the Supplemental Material [23]).

**IV. DISCUSSION**

In the following, we discuss the physical origin of the observed half-integer Shapiro steps in the NiFe JJs. In previous studies, it has been reported that the half-integer Shapiro steps in FM JJs could be induced by two major mechanisms, namely the intrinsic subharmonic CPR effect and the interference of 0- and $\pi$- states [8,9,11,28]. For the intrinsic second harmonic CPR effect, the Josephson critical current can be expressed as $J_C(\theta) = J_{C1}\sin(\theta) + J_{C2}\sin(2\theta)$ ($J_{C1}$ and $J_{C2}$ are the first and second harmonic components) [29]. As the FM thickness increases in the FM JJs, the intrinsic second harmonic component is expected to be much less than the first harmonic component ($J_{C2}/J_{C1} < e^{-d/\xi_{F1}}$, $\xi_{F1}$ is the decaying length in FM) [27]. For NiFe thickness ($d$) of ~ 5.5 nm and $\xi_{F1}$ ~ 1.4 nm [22], $J_{C2}$ is expected to be significant smaller compared to $J_{C1}$ ($J_{C2}$ ~1.8% of $J_{C1}$). Thus, the intrinsic subharmonic CPR effect is believed to play a negligible role in our NiFe JJ. Furthermore, the intrinsic second harmonic CPR induced half-integer steps are expected to be strongly dependent on the temperature [6,26], which does not agree with our experimental observation either (Figs. 4(e) and 4(f)). Furthermore, the nonadiabatic effect driven by the large electromagnetic field has been shown to lead to the fractional order Shapiro steps, such as 1/4, 1/3,1/2, etc. [5,7,30,31], which are temperature dependent. This is not likely to be the major mechanism to account for our observation, since we only observe the half-integer steps and they are robust in all the temperatures from 4 to 7 K. An additional feature of the nonadiabatic effect is that the induced fractional Shapiro steps can be observed above the temperature when the Josephson critical current vanishes [6,7]. However, our experimental results show that the half-integer Shapiro steps disappear as temperature increases when the integer Shapiro steps are no longer observable. All these points present that the nonadiabatic effect does not seem to account for half-integer Shapiro steps in our work. Nevertheless, to fully exclude the nonadiabatic effect, future experimental and theoretical studies will be needed.

To our best understanding, the possible mechanism for the half-integer Shapiro steps will be the interference of 0- and $\pi$- JJs. The superconducting quantum interference device (SQUID) composed of the 0- and $\pi$-junctions, which is called $\pi$- SQUID, is useful to understand the mechanism of interference between the 0- and $\pi$-states [11,33,34]. The $\pi$- SQUID is described by the resistively shunted junction (RSJ) model with two inductively coupled JJs; one is 0- JJ and the other one is $\pi$- JJ. The Josephson coupling energy with respect to the Josephson phases ($\theta$) averaged with two JJs is approximately given by, $\cos(\Psi)\cos(\theta) + \beta \sin^2(\Psi)\cos(2\theta)$, where $\Psi = \pi/2$ for the $\pi$- SQUID and $\Psi = 0$ for the two JJs are both 0- JJs or $\pi$- JJs [34]. The additional phase $\Psi$ due to inductive coupling of 0- and $\pi$- JJs induces the spontaneous circulating current in the $\pi$- SQUID [33]. Here, $\beta$ is given by the inductance ($L$) as $LJ_C/\Phi$. In the $\pi$- SQUID, only the second term, $\beta \sin^2(\Psi)\cos(2\theta)$, remains (Supplemental Material [23]). This means that the interference by $L$ is the origin of the half-integer Shapiro steps.

As well known, the 0-$\pi$ states transition can be induced by varying the FM thickness [21,22,32]. This phenomenon is due to the damping oscillatory behavior of the superconducting order parameter penetrating in the FM layer, where the spatial-dependent wave function of Cooper pairs is always described as $\varphi(d) \propto e^{-d/\xi_{F1}} e^{-id/\xi_{F2}}$ ($\xi_{F2}$ is the oscillation coherent length of superconducting order parameter in FM) [18]. Hence, if there is coexistence of the 0-state and $\pi$-state in NiFe JJs due to spatial variation of the NiFe thickness (see Fig. 5(a)), the half-integer Shapiro steps are possible due to the interference effect between them [33,34].

To investigate this possibility, a high-resolution transmission electrons microscope (HRTEM) has been used to study the NiFe thickness of the strong FM JJ (Figs. 5(b) and 5(d)). Two typical scanning areas A and B, as marked in Fig. 5(b), are characterized, and the results are shown in Figs. 5(c) and 5(d), respectively. Obvious interfaces between Nb and NiFe film can be resolved, as indicated by two dashed orange lines. Subsequently, the NiFe thickness can be obtained to be from ~ 5.0 to ~ 7.0 nm as a function of the position at the interface in area A (Fig. 5(e)). The thickness difference is ~ 2.0 nm, which is about the half-period length of ~ 1.5 nm for the 0-$\pi$ states transition reported in previous NiFe JJs [22]. This indicates that both 0- and $\pi$- states co-exist in our strong FM NiFe JJ. Similar results are observed in area B (Fig. 5(f)). Hence, the experimental results of the NiFe thickness support the scenario that the observed half-integer Shapiro steps are attributed to the co-existence of 0- and $\pi$- states due to the spatial variation of NiFe thickness. The phase diagram of the uniform (0 or $\pi$) region and the coexistence region in the NiFe JJs are studied by numerical simulation (Supplemental material and Fig. S2 [23]). The wide temperature range of the coexistence region can be resolved based on the simplified calculation with $|J_{C0}/J_{C\pi}|$ = 0.9 and 1.1 (Figs. S3(c) and S3(d) in Supplemental Material [23]).

## V. CONCLUSION

In conclusion, we report the experimental observation of the half-integer Shapiro steps in the strong FM NiFe-based JJ. The half-integer Shapiro steps are robust in a wide temperature regime from $T$ = 4 to 7 K, which can be attributed to the co-existence of the 0- and $\pi$- states in the NiFe JJ. This scenario is supported by the characterization of the NiFe thickness variation at the junction using HRTEM. Our results might be important for the FM JJ-based flux qubit applications, and for future investigation of the quantum superposition states [35].

**ACKNOWLEDGEMENTS**

Y.Y., R.C., W.X., Y.M., Y.J., X.C.X., and W.H. also acknowledge the financial support from National Basic Research Programs of China (No. 2019YFA0308401), National Natural Science Foundation of China (No. 11974025), Beijing Natural Science Foundation (No. 1192009), and the Strategic Priority Research Program of the Chinese Academy of Sciences (No. XDB28000000). S.M. is supported by JST CREST Grant (Nos. JPMJCR19J4, JPMJCR1874, and JPMJCR20C1) and JSPS KAKENHI (Nos. 17H02927 and 20H01865) from MEXT, Japan. MM acknowledge the financial support from JSPS (Nos. JP20K03810 and JP21H04987).

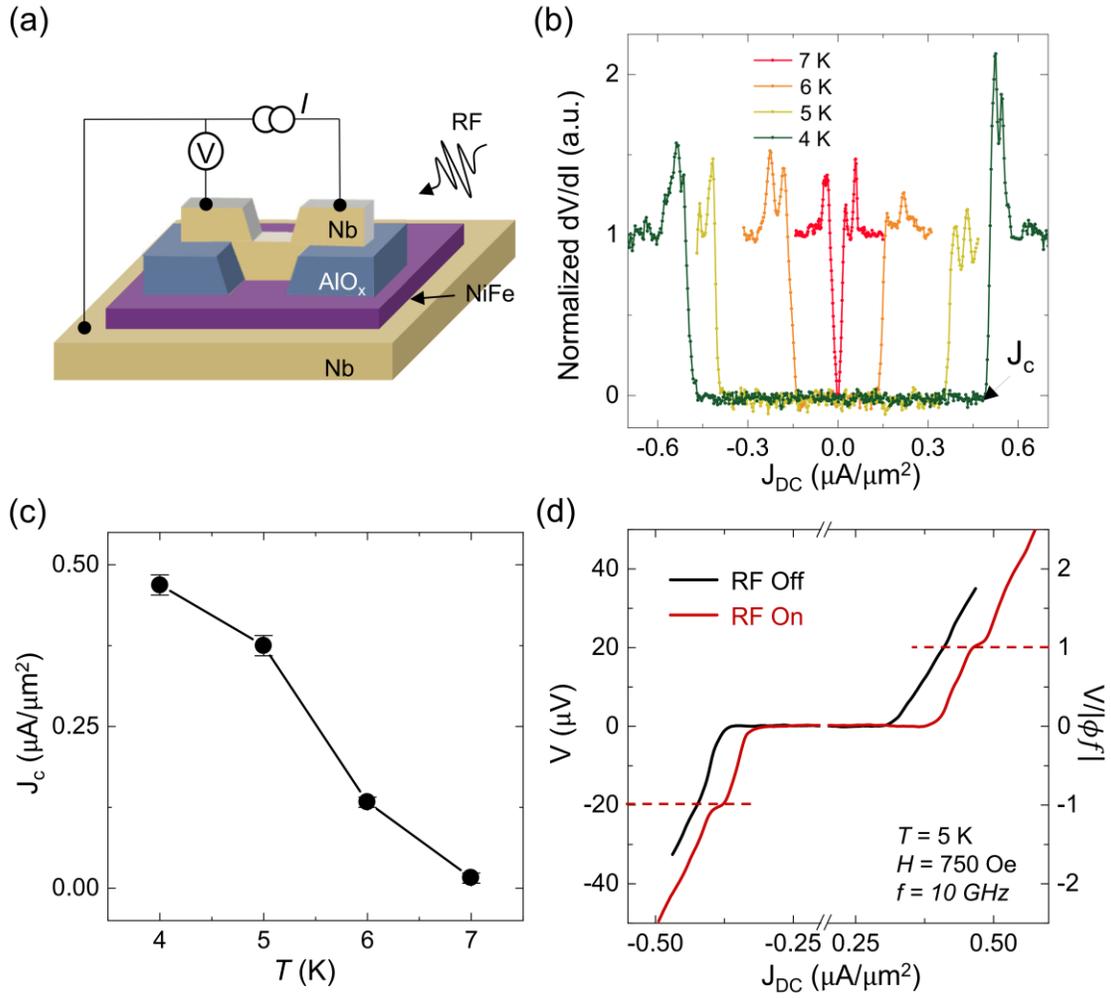

Fig. 1. The basic characteristics of the strong FM NiFe Josephson junction. (a) The schematic of the Nb/NiFe/Nb Josephson device and the measurement geometry. The junction area is ~ 80 μm × 80 μm, and the NiFe layer is ~ 5.5 nm thick. RF represents the radio-frequency ($f$: 1- 20 GHz) microwave excitation using a coplanar waveguide. (b) The normalized differential resistance (dV/dI) as a function of the current density ($J_{DC}$) at $T$ = 4, 5, 6, 7 K, respectively. $J_C$ is the critical current density. (c) The temperature dependence of critical current density of the NiFe Josephson junction. (d) The voltage-current characteristics of the NiFe Josephson junction with and without the microwave excitation ($f$: 10 GHz) at $T$ = 5K and in-plane magnetic field of 750 Oe, respectively. The dashed lines indicate the n = -1 and n = 1 Shapiro steps.

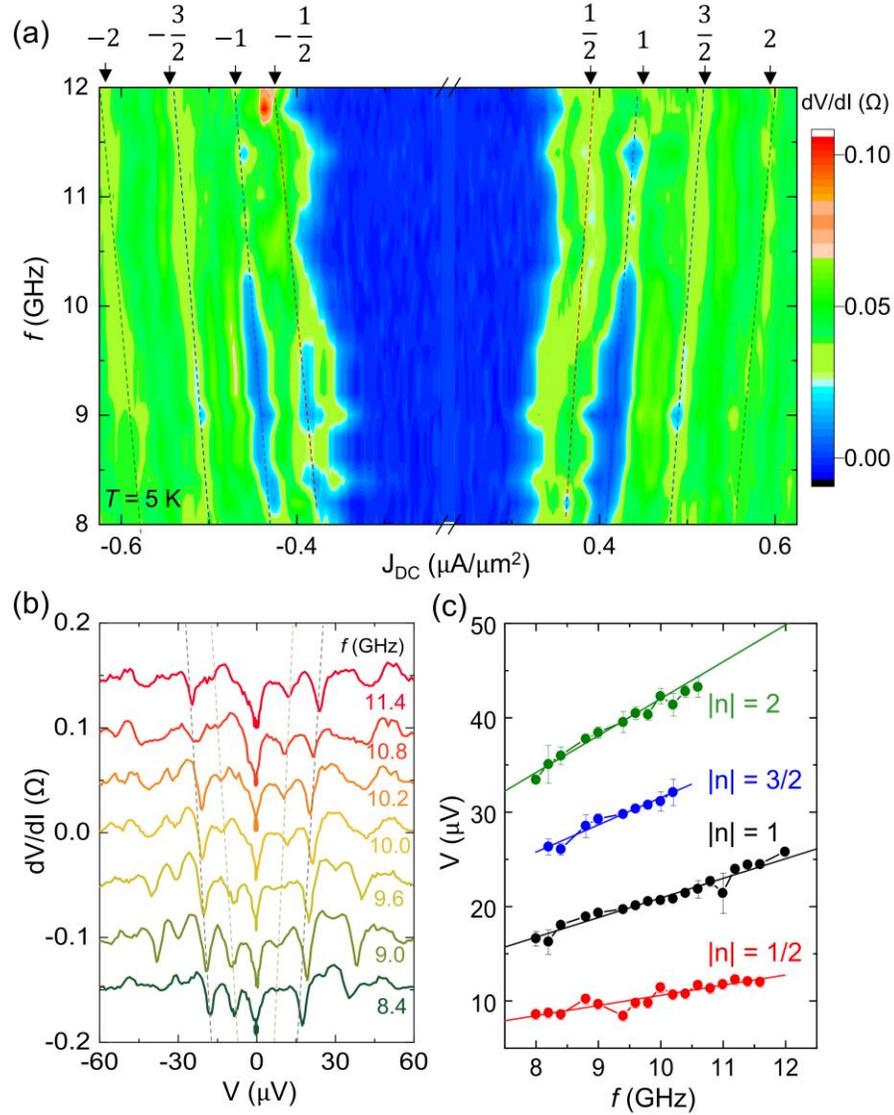

Fig. 2. The half-integer Shapiro steps of the NiFe Josephson device at $T = 5$ K. (a) The contour map of the differential resistance as a function of the current density and RF frequency. The dark blue area represents the superconducting state and the edges are related to the critical current ($J_C$). The dashed lines and the labels indicate the stripes related to the half-integer and integer Shapiro steps. During the measurement, an in-plane magnetic field of 750 Oe was applied. (b) The voltage-dependent differential resistance curves for several representative RF frequencies. These curves are vertically shifted for clarity. The black dashed lines are guides to the eye. (c) The RF frequency dependence of the absolute values of the voltages at the Shapiro steps. The solid lines represent the best linear fitting curves based on the equation of $V = |n\Phi f|$.

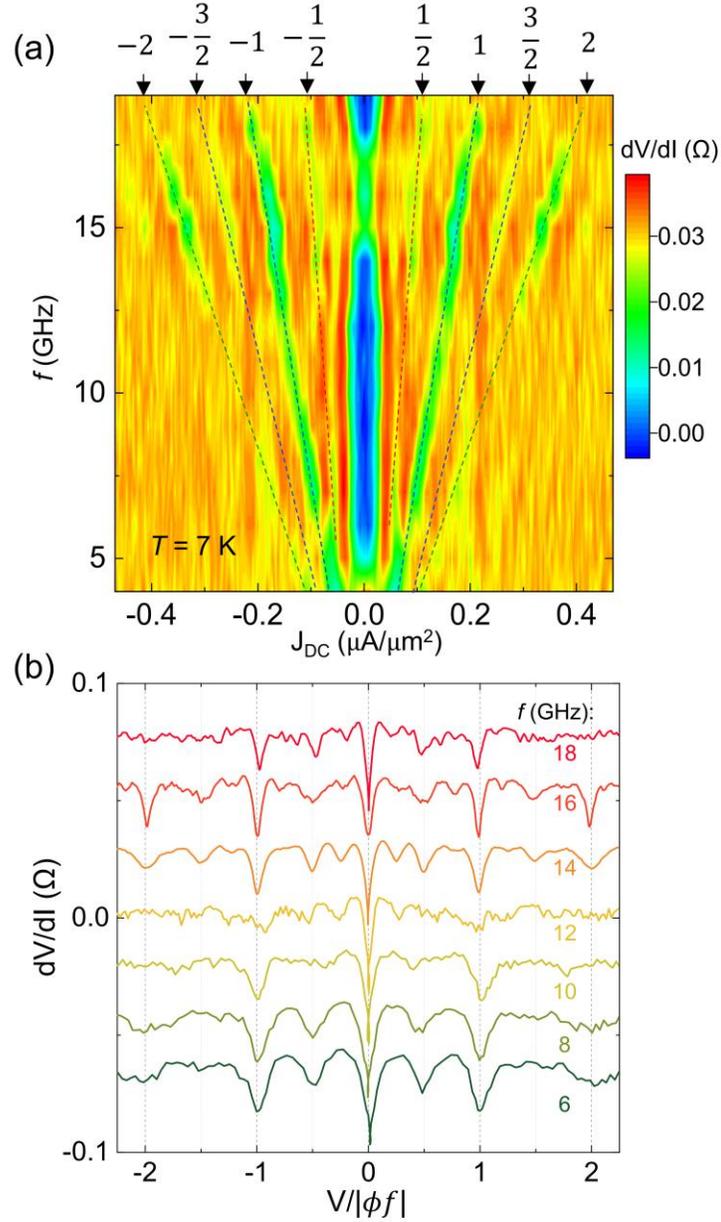

Fig. 3. The half-integer Shapiro steps of the NiFe Josephson device at $T = 7$ K. (a) The contour map of the differential resistance as a function of the current density and RF frequency. The dashed lines and the labels represent the half-integer and integer Shapiro steps. During the measurement, an in-plane magnetic field of 2300 Oe was applied. (b) The differential resistance as a function of the normalized voltage for a series of RF frequencies. These curves are vertically shifted for clarity. The black dashed lines are guides to the eye.

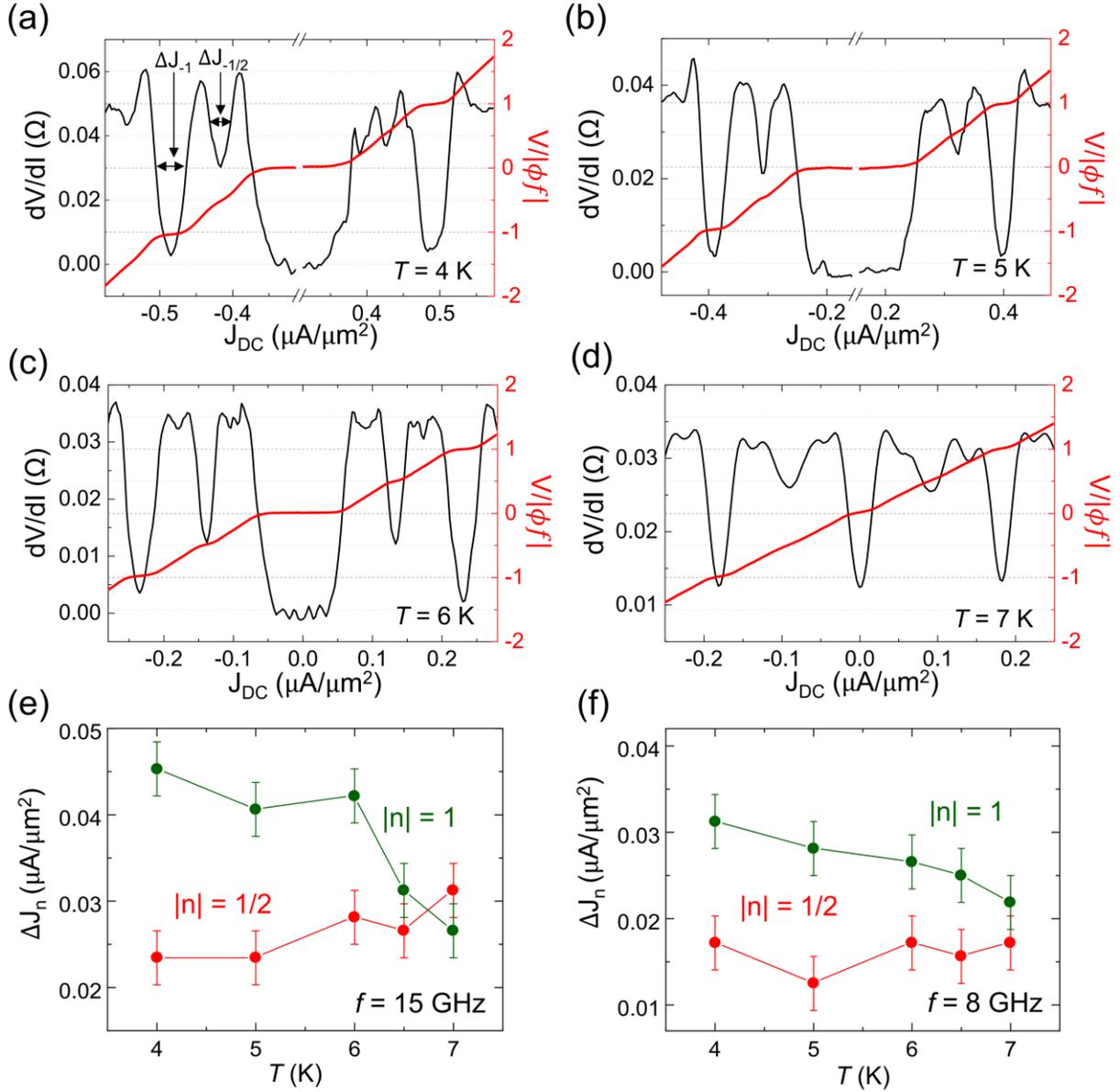

Fig. 4. The temperature dependence of half-integer Shapiro steps. (a-d) The differential resistance (black curves) and normalized voltage (red curves) as a function of the current density at $T$ = 4, 5, 6, and 7 K, respectively. During these measurements, the RF frequency is 15 GHz and the in-plane magnetic field is 2300 Oe. The $\Delta J_{-1}$ and $\Delta J_{-1/2}$ in (a) represent the half-width of the Shapiro steps with index of -1 and -1/2 respectively. (e-f) The temperature-dependent half-width of $|n| = 1$, and $|n| = 1/2$ Shapiro steps for $f$ = 15 and 8 GHz, respectively, where the half-widths are the average values of those with negative and positive index numbers.

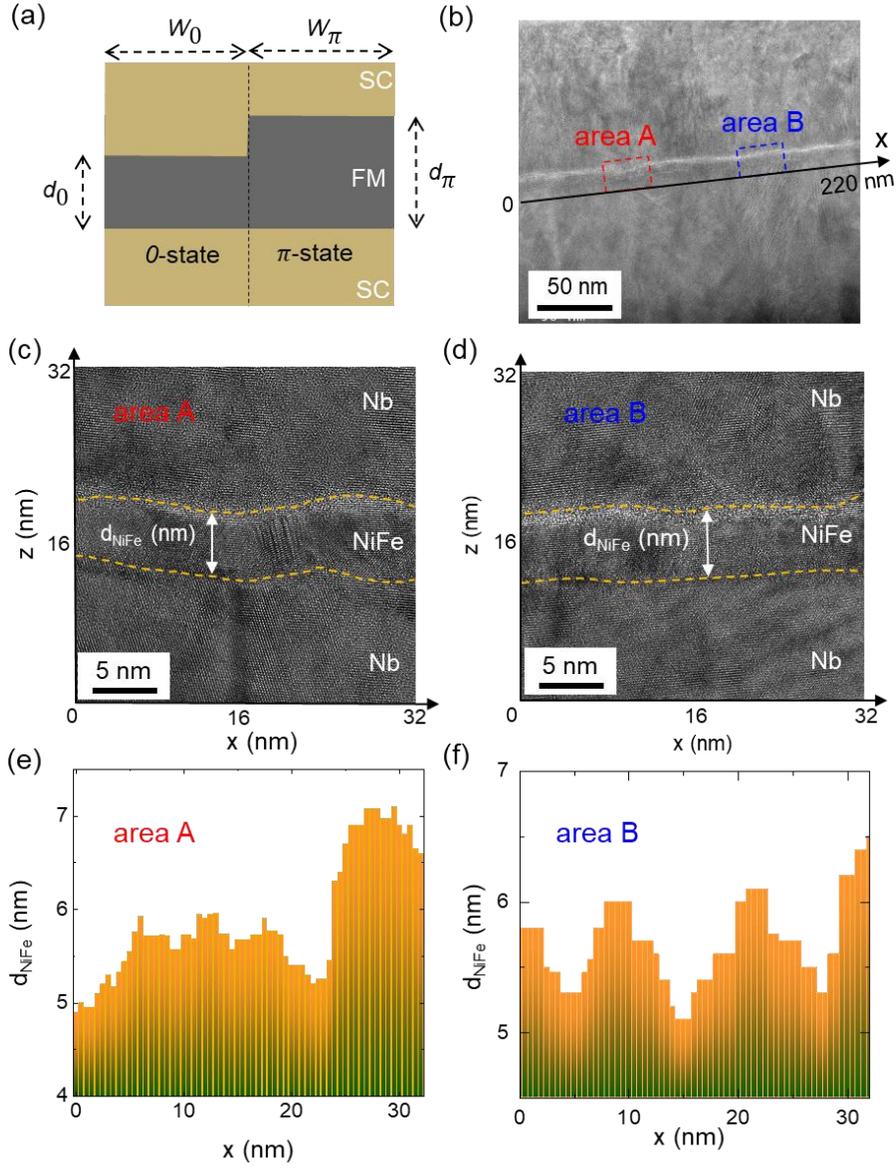

Fig. 5. Analysis of the 0- and $\pi$- states segments in the NiFe Josephson device. (a) The schematic of the coexistence of the 0- and $\pi$- junctions in FM Josephson junction, where $d_\pi$, $d_0$ and $W_\pi$, $W_0$ are the thickness and width of the $\pi$- and 0- junctions, respectively. (b) The coordinate system of two areas A and B for HRTEM characterization. (c-d) The HRTEM characterization of the areas A and B. The dashed lines indicate the two interfaces between NiFe and two Nb layers. (e-f) The spatial distribution of the NiFe thickness along the x axis in Fig. (c) and (d), respectively.

*Supplmentary Materials for:*

# Half-integer Shapiro Steps in Strong Ferromagnetic Josephson Junctions

Yunyan Yao[1], Ranran Cai[1], See-Hun Yang[2], Wenyu Xing[1], Yang Ma[1], Michiyasu Mori[3], Yuan Ji[1], Sadamichi Maekawa[4,5], Xin-Cheng Xie[1,6,7], and Wei Han[1*]

[1] International Center for Quantum Materials, School of Physics, Peking University, Beijing 100871, P. R. China

[2] IBM Research - Almaden, San Jose, California 95120, USA

[3] Advanced Science Research Center, Japan Atomic Energy Agency, Tokai 319-1195, Japan

[4] RIKEN Center for Emergent Matter Science (CEMS), Wako 351-0198, Japan

[5] Kavli Institute for Theoretical Sciences (KITS), University of Chinese Academy of Sciences, Beijing 100049, P. R. China

[6] CAS Center for Excellence in Topological Quantum Computation, University of Chinese Academy of Sciences, Beijing 100190, P. R. China

[7] Beijing Academy of Quantum Information Sciences, Beijing 100193, P. R. China

* Correspondence to: weihan@pku.edu.cn.

## S1. The half-integer Shapiro steps due to coexistence of 0- and π-states.

To understand the mechanism of the interference, the superconducting quantum interference device (SQUID) composed of the 0- and π-junctions, which is called π- SQUID, is useful [1-3]. The π- SQUID is described by the resistively shunted junction (RSJ) model given by,

$$I_j = \frac{V_j}{R} + J_C \sin(\theta_j), \tag{S1}$$

$$\frac{d\theta_j}{dt} = \frac{2\pi V_j}{\Phi_0}, \tag{S2}$$

$$\theta_a - \theta_b = \frac{2\pi\Phi}{\Phi_0}, \tag{S3}$$

$$\Phi = L\frac{I_a - I_b}{2} + \Phi_{ex}, \tag{S4}$$

$$I_a + I_b = I_0 + I_1 \cos(\Omega t) \equiv I_B, \tag{S5}$$

where $j = a, b$ and the flux quantum $\Phi_0 = h/2e$ [1,2]. Each junction is composed of resistance R, Josephson coupling $J_C \sin(\theta_j)$ with current $I_j$, voltage $V_j$, and a phase difference $\theta_j$ at the $j$-junction. The sum of the dc-bias current $I_0$ and the external ac-current $I_1$ with an angular frequency $\Omega$ is the total current, $I_B$, flowing the SQUID. The total flux through the π- SQUID is denoted by $\Phi$ with external flux $\Phi_{ex} = \Phi_0/2$ in our case. Introducing $\theta = \frac{\theta_a + \theta_b}{2}$ and following Ref. [2], the normalized current $i_B = I_B/J_C$ is approximately given by,

$$i_B \sim 2\left[\cos\left(\frac{\pi\Phi_{ex}}{\Phi_0}\right)\sin(\theta) + \beta\frac{1}{8}\sin^2\left(\frac{\pi\Phi_{ex}}{\Phi_0}\right)\sin(2\theta)\right], \tag{S6}$$

Where $\beta = LJ_C/\Phi_0$ with Josephson critical current $J_C$. In the π- SQUID, $\frac{\Phi_{ex}}{\Phi_0} = \frac{1}{2}$ and only the second term, $\beta\frac{1}{8}\sin^2\left(\frac{\pi\Phi_{ex}}{\Phi_0}\right)\sin(2\theta)$, remains. This means that the interference by $L$ is the origin of the half-integer Shapiro steps.

## S2. The estimated phase diagram for the strong FM Josephson device with the coexistence of 0- and π-junctions

Theoretically, for a FM Josephson junction consisting of both 0 and π regions, the total critical current density ($J_C$) can be expressed by the following equation [4],

$$J_C = (w_0 J_{C0} + w_\pi J_{C\pi})/(w_0 + w_\pi), \tag{S7}$$

where $w_0$ and $w_\pi$ are the width of the 0- and $\pi$-junction regions, $J_{C0}$ and $J_{C\pi}$ are the critical currents density for 0- and $\pi$-junctions. The spontaneous circulating currents can lead to the half-integer Shapiro steps (as discussed in Section S1 and equation (S6)), and they depend on the ratio of $w_0$ and $w_\pi$ to their Josephson penetration lengths ($\lambda_{J0}$ and $\lambda_{J\pi}$). The Josephson penetration lengths can be expressed by: $\lambda_{J0} = \sqrt{\hbar/2e\mu_0 d_m J_{C0}}$, and $\lambda_{J\pi} = \sqrt{\hbar/2e\mu_0 d_m J_{C\pi}}$, where $\hbar$ is the reduced Plank constant, $e$ is the electron charge, $\mu_0$ is the vacuum magnetic permeability, and $d_m$ is magnetic barrier thickness ($d_m = 2\lambda + d_{NiFe}$, $\lambda$ is the penetration depth of the superconductor, and $d_{NiFe}$ is the NiFe thickness). The uniform and co-existence phases are separated by the boundary lines defined by $\lambda_{J\pi}\tanh(w_0/\lambda_{J0}) = \lambda_{J0}\tan(w_\pi/\lambda_{J\pi})$ and $\lambda_{J0}\tanh(w_\pi/\lambda_{J\pi}) = \lambda_{J0}\tan(w_0/\lambda_{J0})$ [5,6].

To estimate the phase diagram of the uniform (0 or $\pi$) and 0-$\pi$ co-existence regions in the strong ferromagnetic NiFe JJs, a numerical simulation study is performed. For simplicity, equivalent proportion of junction width ($w_0 = w_\pi = 40\ \mu m$) is assumed. Figure S3 (a-f) show the calculated phase diagram for the uniform and the coexistence region (the yellow area) with $|J_{C0}/J_{C\pi}| = 0.7, 0.8, 0.9, 1.1, 1.2$, and $1.3$, respectively. Next, we calculate the states of the NiFe JJ using these values of $|J_{C0}/J_{C\pi}|$, and $\lambda$ of 100 nm for Nb [7]. These calculated results at $T = 4, 5, 6, 7$ K are represented by the red stars in Fig. S3(a-f). Clearly, the existence of the half-integer Shapiro steps is robust from $T = 4$ to 7 K, and all of them could be resolved in the numerical simulation with $|J_{C0}/J_{C\pi}| = 0.9$ and 1.1 (Fig. S3(c-d)).

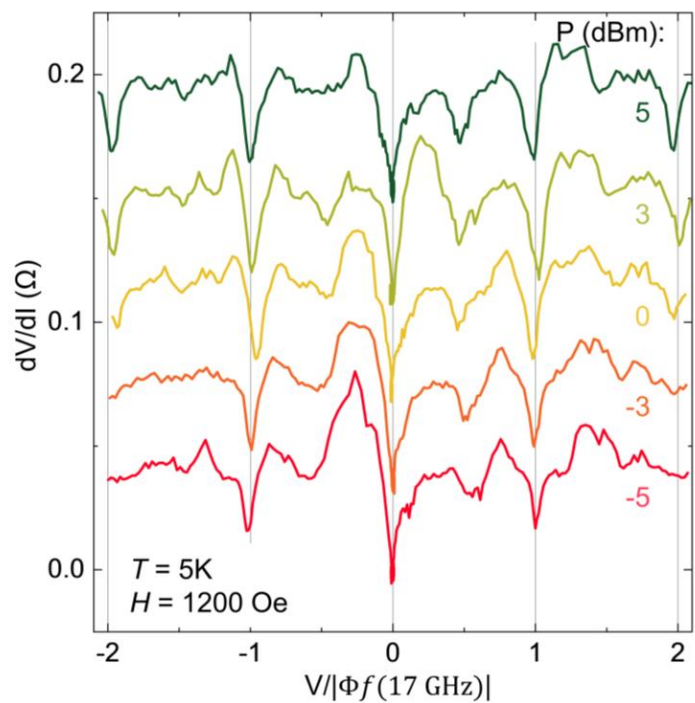

Fig. S1. The RF power dependence of the Shapiro steps at $T$ = 5 K and under $H$ = 1200 Oe. These cures have been vertically shifted for clarity.

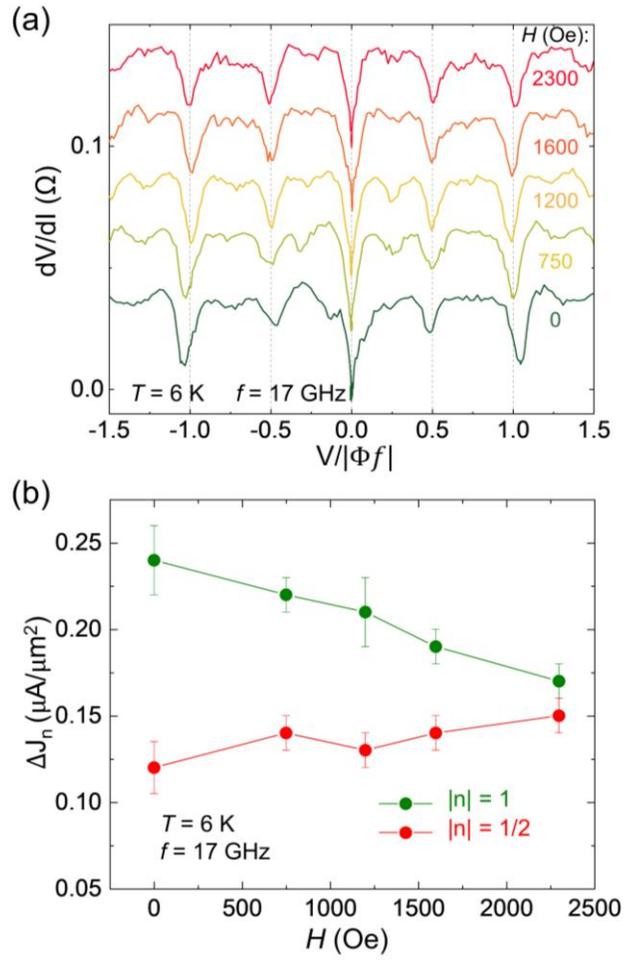

Fig. S2. The in-plane external magnetic field dependence of the Shapiro steps at $T$ = 6 K. (a) The differential resistance vs. the normalized voltages for the magnetic field ranging from 0 to 2300 Oe. These cures have been vertically shifted for clarity. (b) The magnetic field dependence of the step width of the integer (green) and half-integer steps (red).

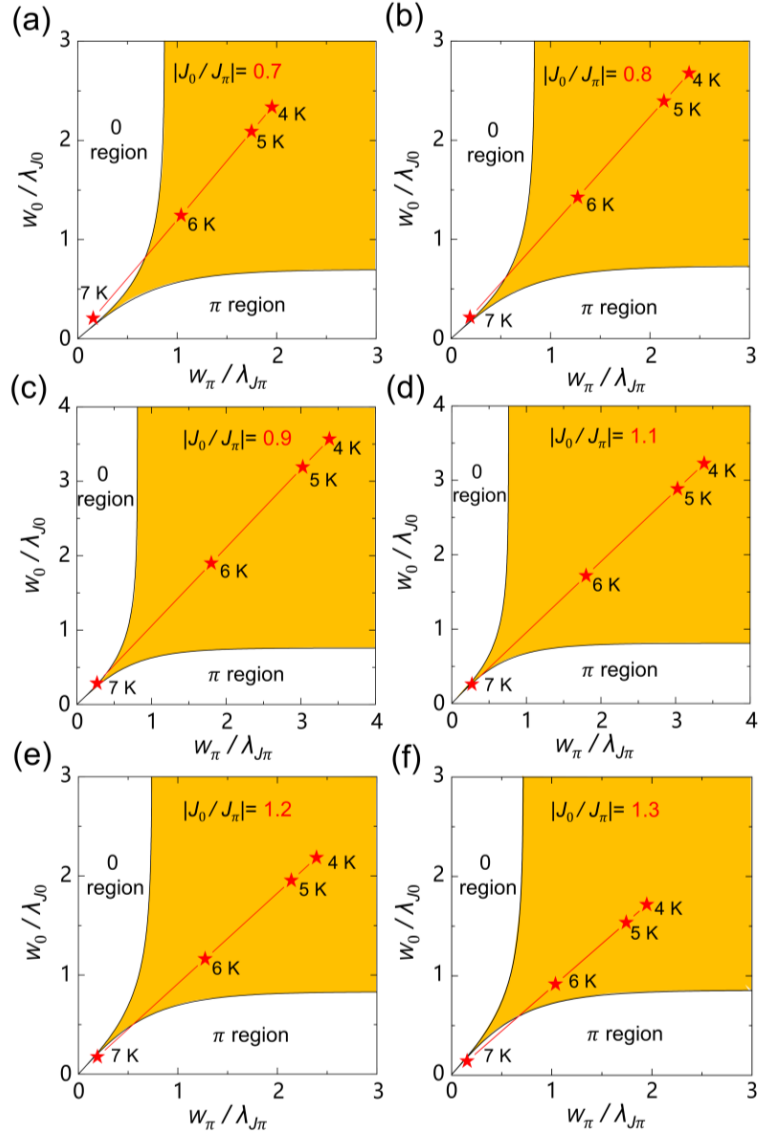

Fig. S3. The estimated phase diagram of the uniform (0 or $\pi$) and the 0-$\pi$ co-existence regions. (a-f) The estimated phase diagram and the estimated co-existence states for our experimental results (red stars, $T$ = 4, 5, 6, and 7 K). The numerical simulation is performed under $w_0/w_\pi = 1$ and $|J_{C0}/J_{C\pi}|$ = 0.7, 0.8, 0.9, 1.1, 1.2, and 1.3, respectively.

23